\documentclass[10pt, twocolumn]{article}
\usepackage{needspace}
\usepackage{balance}
\usepackage{hyperref}
\usepackage{hhline}
\usepackage{multirow}
\usepackage{adjustbox}

\begin{document}

\title{Mutation Testing as a Safety Net \\ for Test Code Refactoring}

\author{
Ali Parsai, Alessandro Murgia, Quinten David Soetens, and Serge Demeyer \\
Antwerp Systems \& Software Modelling (Ansymo)\\
University of Antwerp\\
Middelheimlaan 1\\
2020 Antwerp, Belgium\\
ali.parsai@student.uantwerpen.be,\\ \{alessandro.murgia; quinten.soetens; serge.demeyer\}@uantwerpen.be
} 
\date{}

\maketitle
\begin{abstract}
Refactoring is an activity that improves the internal structure of the code without altering its external behavior. When performed on the production code, the tests can be used to verify that the external behavior of the production code is preserved. However, when the refactoring is performed on test code, there is no safety net that assures that the external behavior of the test code is preserved.\\
In this paper, we propose to adopt mutation testing as a means to verify if the behavior of the test code is preserved after refactoring. Moreover, we also show how this approach can be used to identify the part of the test code which is improperly refactored. 
\end{abstract}

\section{Introduction}
\label{Introduction} 

Refactoring is ``the process of changing a software system in such a way that it does not alter the external behavior of the code yet improves its internal structure''~\cite{Fowler1999}.
If applied correctly, refactoring improves the design of software, makes software easier to understand, helps to find faults, and helps to develop a program faster~\cite{Fowler1999}. However, the process of refactoring is not always performed flawlessly~\cite{Kim2011,Soares2010}, leading to faults being introduced into the refactored code due to mistakes made by developers, or using automated refactoring tools that do not preserve the behavior of the code. Thus, there is a need for a safety net that saves developers when they refactor improperly the test code.

Refactoring does not only target the production code, but it also actively involves the test code. 
Ideally, in object-oriented systems, for each production class, we have a related counterpart in the test section. 
As a consequence, the size of the test suite increases linearly with the size of the system. 
This scenario is particularly common in software systems developed using test-driven development, since it leads
to a rapid development of test suites~\cite{Beck2002}. 
In this context, it is important to also refactor the test code to keep it synchronized with the evolution of the production code and avoid its quality erosion~\cite{Rompaey2006}. 

Refactoring of the production code can be done with less risks using a test suite, 
since it provides a safeguard against regressions during software transformation~\cite{Vonken2012}. 
Tests ensure that the production code preserves its external behavior pre- and post- refactoring.
On the contrary, there is no widely-accepted %
method to verify if a refactored test suite preserves its external behavior. 
Several studies point out the peculiarities of test code refactoring~\cite{Deursen2001,vanDeursen2002,Counsell2006,Pipka2002}. However, none of them provided an operative method to guarantee that such refactoring was preserving the behavior of the test.
To address this shortcoming, we propose the adoption of mutation testing as a safety net for test code refactoring.

Mutation testing provides a repeatable and scientific approach to measure the quality of the test code. It consists of two phases: First, generating faulty versions of the code by injecting a single fault into the code \textit{(creating a mutant)} and then, executing the test suite on this faulty version of the code to determine the outcome. The output of mutation testing is a quality metric (mutation coverage) defined by the percentage of the faults that resulted in failure of at least one test \textit{(killed mutants)} by the total number of created mutants. This metric is proven to simulate the faults realistically~\cite{Andrews2005,Just2014}. This is due to the fact that the faults introduced by each mutant are modeled after the common mistakes developers often make~\cite{Jia2011}.

A correctly performed refactoring of the test code should not change its external behavior, and consequently its mutation coverage should remain unaltered. For this reason, we propose to calculate mutation coverage of each class in the production code pre- and post- refactoring of the test code. The comparison between both reveals whether the refactoring had any effect on the tests covering that class. 
Moreover, this approach points out the location of the injected faults helping to spot easily which part of the test code was improperly refactored.

To validate our approach we run the experiments on two projects. The first project is a simple system created \textit{ad hoc} to show how mutation testing is capable of identifying a change in the external behavior of a refactored test. The second project is used to verify our approach in a real open source system with a refactored test suite.

The paper has the following structure.
Section \ref{Background} reports the background notions related to mutation testing.
Section \ref{ExperimentalSetup}  describes the research approach adopted.
Section \ref{Results}, discusses the results of our research.
Section \ref{ThreatsToValidity} presents the threats to validity. 
Section \ref{RelatedWork} reports the related work.
Finally, section \ref{Conclusion}  summarizes our findings.

\section{Background}
\label{Background}
This section describes the typical quality metrics of the test code, and background information related to mutation testing. In addition, it provides an overview of the implementation of mutation testing in LittleDarwin\footnote{http://littledarwin.parsai.net/}.

\subsection{Simple Coverage Metrics}
There are simple coverage metrics available to estimate the quality of a test suite~\cite{Zhu1997}. \textit{Statement coverage} determines the percentage of executed statements by test code. In a similar fashion, \textit{Branch coverage} determines the percentage of the branches of code that are executed by the test code. A branch is created in a program when a control statement (e.g. \textit{if} or \textit{switch} statements) provides two or more paths of execution. These metrics provide an overview of the quality of the test suite in an easily attainable manner; Yet, they are inadequate in their purpose of estimating quality~\cite{Wei2012}. Even a 100\% branch coverage would leave a lot of room for a fault to escape~\cite{Marick1991}. Furthermore, branch coverage is also a poor measure to determine a detailed map of the weaknesses in a test suite because first, it lacks the ability to discover which type of faults are being caught, and which are not; and second, it is difficult for practical tools to trace the execution paths during the runtime of complicated software systems. Thus, these metrics are not adequate enough to discover small mistakes in the test code, and to trace back the change in behavior to the faulty code.

\subsection{Mutation Testing}

\begin{figure}
	\centering
	\includegraphics[width=0.7\linewidth]{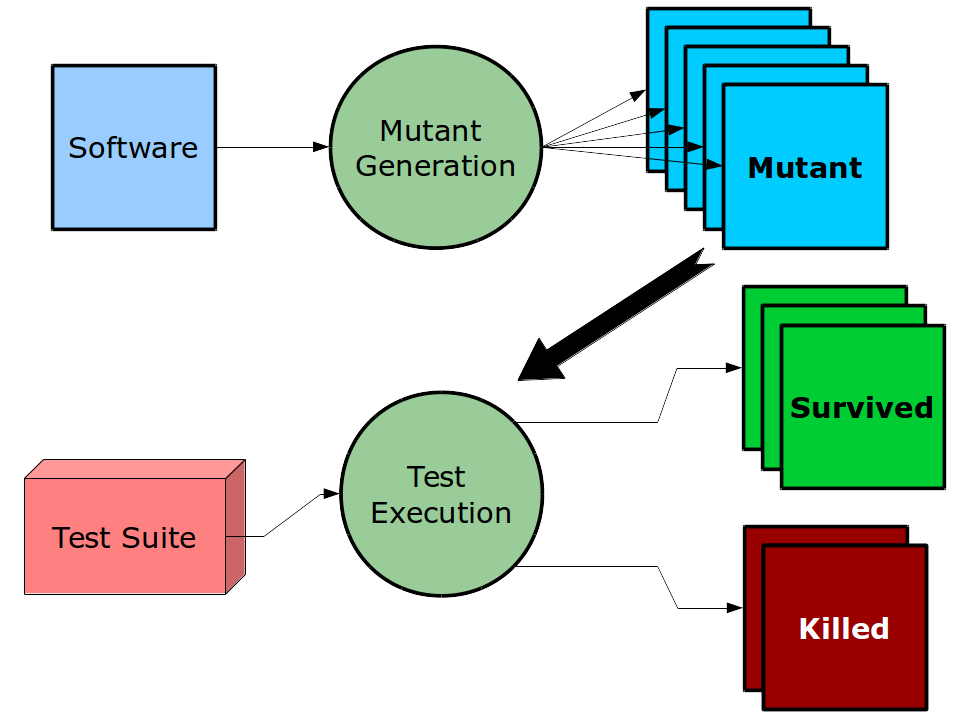}
	\caption{Mutation testing procedure}
	\label{fig:mutationanalysis}
\end{figure}

Mutation testing is the process of injecting faults into software, and counting the number of intentional faults which make at least one test fail. The idea of mutation testing was first mentioned in a class paper by Lipton~\cite{Offutt2001} and later developed by DeMillo, Lipton and Sayward~\cite{DeMillo1978}. The first implementation of a mutation testing tool was done by Timothy Budd in 1980~\cite{Budd1980}. This procedure is executed in the following manner: First, faulty versions of the software are created by introducing a single fault into the system \textit{(Mutation)}. This is done by applying a known transformation on a certain part of the code \textit{(Mutation Operator or Mutator)}. The more mutants generated for a class, the more chance that we detect a change in behavior.  After generating the faulty versions of the software \textit{(Mutants)}, the test suite is executed on each one of these mutants. If there is an error or failure during the execution of the test suite, the mutant is regarded as \emph{killed}. On the other hand, if all tests pass, it means that the test suite could not catch the fault and the mutant has \emph{survived}. This procedure demands a \textit{green} test suite ---a test suite in which all the tests pass--- to run correctly. An overview of this procedure can be observed in Figure~\ref{fig:mutationanalysis}. 

\begin{equation}
\label{coverageequation}
Mutation\ Coverage = \frac{Killed\ Mutants}{All\ Mutants}
\end{equation}

The final result is calculated using Equation~\ref{coverageequation}. This metric provides a more detailed image of the quality of a test by emphasizing test results. This makes sure that the kind of faults simulated by mutation operators are covered by the test; Therefore reducing the chance of missing such faults in the final product.

\subsection{LittleDarwin}
LittleDarwin\footnotemark[1] is a mutation testing tool created by the first author. This tool is designed to offer an alternative framework for those who need to apply mutation testing to complex systems; but it is capable of analyzing simple systems as well. LittleDarwin is designed to be independent from the testing structure. As a result, LittleDarwin demands much less compatibility from the target system in order to perform its analysis. Thus, it can be run on any build structure no matter how complex it is, given following conditions:

\begin{enumerate}

	\item The build process must be able to run the test suite.
	\item The build process must return non-zero if any tests fail, and zero if it succeeds.
	\item The build process must be sufficiently fast in order to keep the total run time practical.

\end{enumerate}

LittleDarwin is designed in an expandable way, so that interested developers can develop their own mutation components and still use the structure of the main software to run the mutation analysis. This broadens the scope of its applicability. In its current version, LittleDarwin supports mutation testing of Java programs.

In total, there are 9 mutation operators implemented in LittleDarwin. %
These mutators are collectively known as the minimal set. The description of each mutator along with an example can be found in Table~\ref{mutationoperators}.

\begin{center}
	\begin{table}
		\centering
		\adjustbox{max width=\columnwidth}
		{\begin{tabular}{|l||l|c|c|}
			\hline \multirow{2}{*}{\textbf{Operator}} & \multirow{2}{*}{\textbf{Description}} & \multicolumn{2}{c|}{\textbf{Example}} \\
			\hhline{~~--} & & \textbf{Before} & \textbf{After} \\ 
			\hline
			\hline AOR-B & Replaces a binary arithmetic operator & $a + b$  & $a - b$ \\ 
			\hline AOR-S & Replaces a shortcut arithmetic operator & $++a$ & $--a$ \\ 
			\hline AOR-U & Replaces a unary arithmetic operator & $-a$ & $+a$ \\ 
			\hline LOR & Replaces a logical operator & $a\,\&\,b$ & $a\,|\,b$ \\ 
			\hline SOR & Replaces a shift operator & $a >> b$ & $a << b$ \\ 
			\hline ROR & Replaces a relational operator & $a >= b$ & $a < b$ \\ 
			\hline COR & Replaces a binary conditional operator & $a\:\&\&\:b$ & $a\,||\,b$ \\ 
			\hline COD & Removes a unary conditional operator & $!\,a$  & $a$ \\ 
			\hline SAOR & Replaces a shortcut assignment operator & $a\:*= b$ & $a\:/= b$ \\ 
			\hline 
		\end{tabular}}
		
		\caption{LittleDarwin mutation operators}
		\label{mutationoperators}
	\end{table}
\end{center}

\section{Experimental Setup}
\label{ExperimentalSetup}
As described before, the problem we try to solve is to detect improper refactoring. So, we aim to provide an operative method to verify whether or not the refactoring activity has changed the external behavior of the test suite. To achieve this, we examine two cases:

\begin{itemize}
\item A use case where an 'improper'  refactoring of the test code \textit{changes} the testing behavior.
\item A use case where a 'proper' refactoring of the test code \textit{does not} change the testing behavior. 
\end{itemize}

Two projects are selected as our cases. First, we create a toy project to exhibit the ability of mutation testing  to highlight a change in the behavior of the test suite. We use the same project to demonstrate the usage of mutation testing to identify the improperly refactored part of the test code.  
Second, we use an open source project %
to verify how our approach can be applied on real refactorings that affect the test code.

Each project has two versions: pre- and post- refactoring. 
For each version, we use JaCoCo\footnote{http://www.eclemma.org/jacoco/} to calculate statement and branch coverage, and LittleDarwin to run mutation testing.%
\needspace{5\baselineskip}

\subsection{Toy Project}

\begin{figure}
	\centering
	\includegraphics[width=\linewidth]{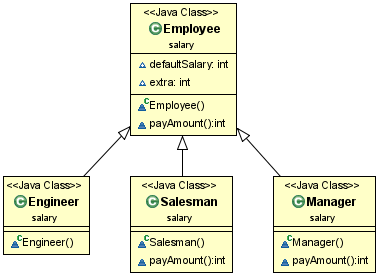}
	\caption{UML class diagram of the toy project}
	\label{fig:toyprojectstructure}
\end{figure}

\begin{figure*}
		\centering
		\includegraphics[width=\textwidth]{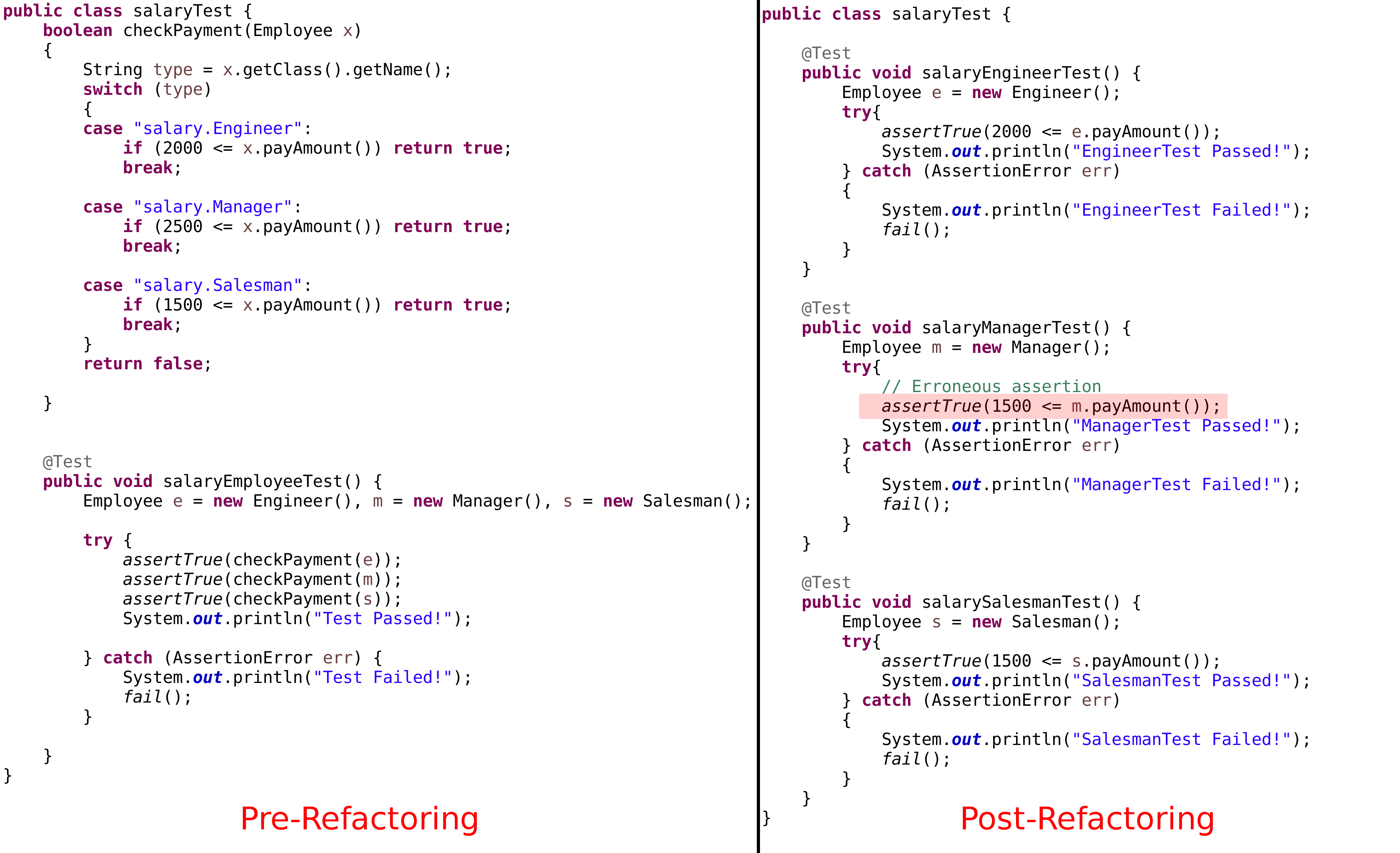}
		\caption{Test class pre- and post- refactoring}
		\label{fig:testclassrefactoring}
	\end{figure*}

We designed a simple project  inspired by the example proposed by Fowler, et al.\cite[p. 207]{Fowler1999}.
In Figure~\ref{fig:toyprojectstructure} we show the UML class diagram related to the production code of this project. 
In Figure~\ref{fig:testclassrefactoring} we report the two versions of the test code: pre- and post- refactoring. 

The test pre- refactoring suffers
from two code smells (1) conditional statement that checks the type of the input variable~\cite{Fowler1999}, and (2) assertion roulette~\cite{Moonen2008}. 
In the test post- refactoring these code smells are removed by introducing three separate test methods.
Here, we simulate the introduction of a naive mistake (Figure~\ref{fig:testclassrefactoring}, red area): during the operation of copy \& paste the developer did not correctly adapt the method \texttt{salaryManagerTest()}. In the post- refactoring version, instead of the correct value of \textit{2500}, the value is set to be \textit{1500}. 
This mistake is introduced to show (in Section \ref{Results}) how mutation testing can detect behavior change and be used to trace back improperly refactored tests.

	\begin{figure}
		\centering
		\includegraphics[width=\linewidth]{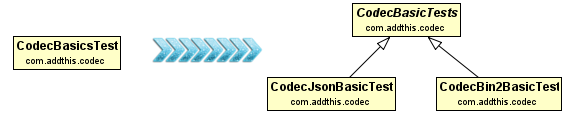}
		\caption{Extract Class refactoring in the test code of Codec}
		\label{fig:codecrefactoring2}
	\end{figure}

\subsection{Real Project}
We analyze the open source project Codec\footnote{http://github.com/addthis/codec/}.
Using Ref-Finder~\cite{Kim2010}, we identify which refactorings were performed during its evolution and among them which 
ones affected the test code. All of them were manually validated. Moreover, during the manual inspection few other refactorings were 
added to the list.

Table~\ref{refactorings} reports all test refactorings identified. As we can see \texttt{CodecBasicsTest} is involved in several types of refactorings. 
One of the relevant refactorings in terms of test reengineering was the extraction of several nested classes from \texttt{CodeBasicTest}, which
was later on followed by a further extract class refactoring  (Figure~\ref{fig:codecrefactoring2}). In this refactoring, two extra test classes were created to incorporate assertions related to their corresponding classes in the production code, thus eliminating assertion roulette and god class code smells in \texttt{CodecBasicsTest}.

In the toy project the refactoring does not affect the production code. 
For this reason we were able to verify on the \textit{same} version of the production code whether the refactoring of the test code preserved its external behavior. 
In a real project like Codec, refactorings and other maintenance activities co-occur in production and test code. 
In this scenario, we cannot verify if the refactored test suite is changing its behavior due to production code change.
For this reason, we had to introduce an alternative version of Codec in our experiment, in which the refactoring is restricted to the test code. 
To accomplish this task, we had to go through the history of the project and:
 \begin{enumerate}
\item Identify when a refactoring is performed on the test suite. 
\item Back port this refactoring to the previous version of system
\item Create an alternative version of the system where the production code is the same, but the test code is refactored. We call this the post-clean- refactoring version.
 \end{enumerate} 
 
At the end of the process, we have two versions of the system: pre- refactoring and post-clean- refactoring. Both versions have the same production code, but differ in the refactoring of the test suite.
These two versions are the ones we use to verify whether the test refactoring modifies its external behavior.

\begin{center}
	\begin{table}
		\centering
		\adjustbox{max width=\columnwidth}
{
	\begin{tabular}{|l|l|l|}

		\hline Refactoring & Target Class & Instances\\ 
		\hline
		\hline Remove Parameter &  CodecGenericsTest & 1\\ 
		\hline Add Parameter & CodecGenericsTest & 1\\ 
		\hline Rename Method & CodecRWOnlyTest & 1 \\ 
		\hline Move Method & CodecBasicsTest & 1 \\ 
		\hline Extract Nested Class & CodecBasicsTest & 12   \\ 
		\hline Extract Class & CodecBasicsTest & 1 \\
		\hline Rename Class & CodecBasicsTest & 1 \\
		\hline Remove Control Flag & CodecBasicsTest & 1 \\ 
		\hline Replace Magic Number with Constant & CodecTest & 2  \\ 
		\hline Replace Magic Number with Constant & CodecUtilTest & 1  \\ 
		\hline Replace Magic Number with Constant & CodecObjectSubclassTest & 1  \\ 
		\hline
		
	\end{tabular}}			
			\caption{Refactorings in Codec test code}
			\label{refactorings}
		\end{table}
	\end{center}

\section{Results}
\label{Results} 

In this section we show how mutation coverage is able to highlight whether or not a test refactoring causes a modification of its 
 external behavior. For the toy project we compute mutation coverage along with percentage of passed tests, statement and branch coverage to
prove that these approaches are not suitable for detecting a change in the test  behavior. For the Codec project, we limit our analysis to mutation coverage.

\subsection{Toy Project}
The toy project has a refactoring of the test code. 
One of the refactorings was improperly done: a fault was introduced in \texttt{salaryManagerTest}.
Here we compute the three metrics that are generally used to evaluated test quality: 
percentage of passing tests, statement and branch coverage and
mutation coverage.
The results are presented in figure~\ref{fig:toyexampleresults}. Here we can see:

	\begin{figure}
		\centering
		\includegraphics[width=\columnwidth]{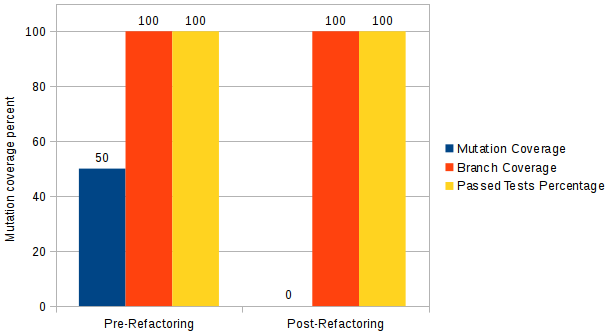}
		\caption{Percentage of passing tests, statement and branch coverage and
mutation coverage for all classes of the toy project}
		\label{fig:toyexampleresults}
	\end{figure}	
	
\begin{figure}
	\centering
	\includegraphics[width=0.8\columnwidth]{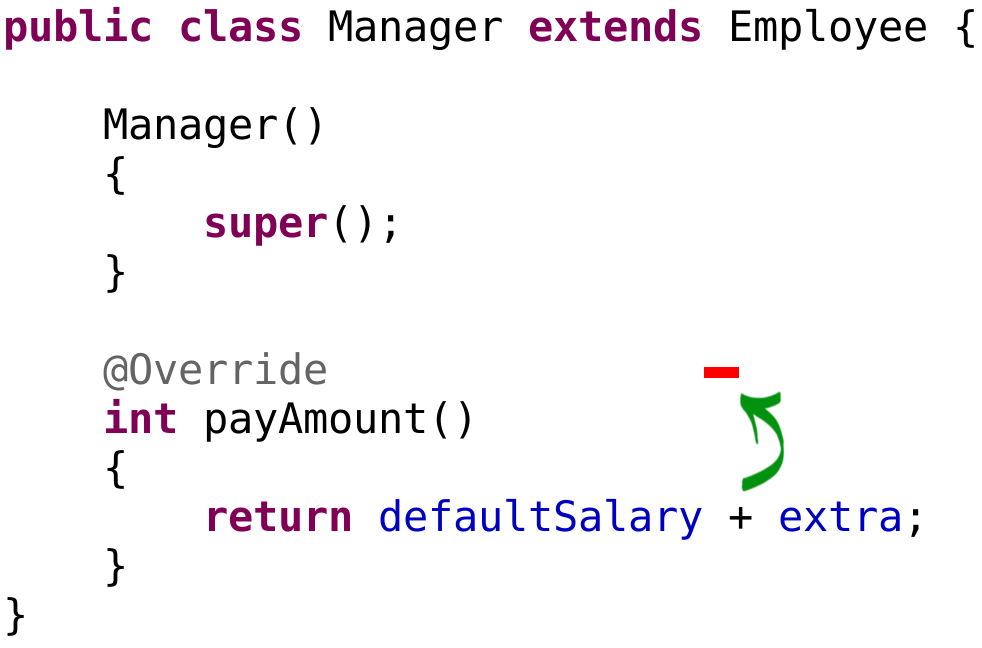}
	\caption{The mutant that changes the status pre- and post- refactoring in the toy project}
	\label{fig:toyexamplemutant}
\end{figure}

\begin{itemize}
	\item \textbf{Percentage of passing tests.} In both pre- refactoring and post- refactoring versions, all tests passed. The wrongly refactored method \texttt{salaryManagerTest} was not detected.
	
	\item \textbf{Statement, and branch coverage.} Both metrics grant a 100\% coverage for the pre- and post- refactoring versions.  
	Also in this case \texttt{salaryManagerTest} was not detected as a faulty refactored test.

	\item \textbf{Mutation coverage.} For the mutation analysis, LittleDarwin introduced two mutants in the production code by
	replacing the operators  \texttt{+} with  \texttt{-} and vice-versa (figure~\ref{fig:toyexamplemutant}).
	In the pre- refactoring version, one of these mutants was killed, resulting in a 50\% mutation coverage. Whereas, 
	in the post- refactoring version both mutants survived, resulting in a 0\% mutation coverage. 
	The different mutation coverage is the first hint that the refactoring \textit{changed} the external behavior of the test code.
	
	Investigating on which mutant changed the status of the test (passing to not passing or vice versa), 
	we trace  the problem back to \texttt{salaryManagerTest}. Finally, comparing the two versions of the test code we identify the fault\footnote{ 
      The manual analysis of two version of system  %
	 is practical for small projects where the developer has a complete understanding of the code. In larger projects, it can be performed automatically using a dynamic analysis tool to find out the relationship between the tests and the classes of the production code.}.	

\end{itemize}

\subsection{Real Project}
\begin{figure}
	\centering
	\includegraphics[width=\columnwidth]{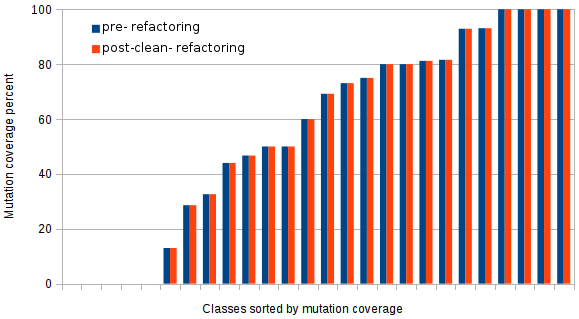}
	\caption{Mutation coverage for each class of the Codec project}
	\label{fig:realprojectresults}
\end{figure}

  The Codec project presents realistic refactorings applied to the test suite.  In this project we do not compute 
 the metrics number of passing tests, statement and branch coverage since these were were inadequate to highlight a change of behavior due to test refactoring. For the Codec project, by only computing the mutation coverage, we obtain the results in figure~\ref{fig:realprojectresults}.
 As we can see the number of mutants killed (or survived) in pre- and post-clean- refactoring is the same. This implies that all refactorings, 
 including the major ones, were properly performed, since they did not change the external behavior of the test code.

\section{Threats To Validity}
\label{ThreatsToValidity}

In this section we present the threats to validity of our study according to the guidelines reported in~\cite{Yin2003}. 

Threats to internal validity concern confounding factors that can influence the obtained results. In this study,
 the mutation coverage %
 depends on the set of mutation operators used in mutation testing process. As a consequence, the ability of mutation testing in 
 detecting  behavioral changes is limited by the type of operators adopted.
To alleviate this problem, we used the standard set of mutation operators that most mutation testing tools support~\cite{Parsai2015} since they are
able to reflect the mistakes commonly introduced by developers. In case of Object-Oriented Programming languages, an additional set of mutation operators were designed by Ma et al.~\cite{Ma2002} to model mistakes specific to these languages. Using these operators would lead to more accurate results by introducing the support for new types of mistakes. This increased accuracy might be necessary in case of software systems that make use of Object-Oriented Programming structures widely throughout the code.
Another threat stems from the masking effect provided by multiple failings tests. %
This happens when a mutant is killed in the original version with the failure of some tests, and the behavioral change causes another test to fail on the same mutant. In this case, the mutant would not change status, and therefore, the mutation coverage would stay the same. 
Solving this problem is not trivial 
 and requires more research on the subject.   

Threats to construct validity focus on how accurately the observations describe the phenomena of interest.
 For our experiment, the elements of interest are (1) the ability of mutation testing in verifying whether test code refactoring was improperly performed and in that case (2) which section of the test code is the cause. 
We used a real project as well as our toy project to cover element (1), while the experiment on the toy project also contained element (2). 
Even though the toy project is very small, its improper refactoring is still valuable since representative of a common mistake. 

Threats to external validity correspond to the generalizability of our results. In this experiment we use only two projects.
Although one of them was representative of a real open source project, actively maintained and developed by a commercial company, it is desirable to replicate this study taking into more projects; especially the ones where the test code was modified with refactorings different from the one we considered.

Threats to reliability validity correspond to the degree to which the result depends on the used tools. We depend on Ref-Finder and manual inspection to discover the refactoring of the test code in our real project case. There is a possibility that we miss some refactorings or make mistakes in this process. 
We counter this chance  
by checking our list of refactorings against the code changes between two versions. 
We also depend on the tools JaCoCo (to calculate statement and branch coverage) and LittleDarwin (to calculate mutation coverage). 
The outcome of JaCoCo has been manually verified due to the simple nature of our toy project. The outcome of LittleDarwin has been tested and explored  
in the first author's masters thesis~\cite{Parsai2015}.

\section{Related Work}

Our study refers to the adoption of mutation testing in the context of test refactoring.
For this reason we present the related work divided in two parts.

\label{RelatedWork}

\subsection{Mutation Testing}
One of the articles that performs a comprehensive analysis of the subject is Jia et al. 2011~\cite{Jia2011}. This article is a literature survey that tries to summarize a huge amount of information about the process of mutation testing, performance, practicality, etc. 
Offutt et al. in~\cite{Offutt2011} discuss the history of mutation testing and the state of the art, and provides insight into the future of the field.
A good reference for analysis of the mutation testing tools for Java is Delahaye et al. 2013 ~\cite{Delahaye2013}. 
In the mentioned article, mutation testing tools for Java are compared based on efficiency, compatibility with current technologies and multiple other factors.

Previous studies discuss mutation testing from different point of views. However, none of them 
propose the adoption of mutation testing to analyze the behavior preservation in the context of test refactoring. 

\subsection{Test Refactoring}

The concept of refactoring and behavior preservation was introduced for the first time by  Opdyke \cite{Opdyke1992}. 
However, this work does not differentiate between refactoring of the test code and refactoring of production code. 
Later on, several studies discovered and investigated the peculiarity test code refactoring.
van Deursen et al. were the first to highlight the characteristics of test refactoring by providing
a list of test code smells and test-oriented refactorings \cite{Deursen2001}. van Deursen and Moonen identified how refactoring can affect the test code \cite{vanDeursen2002}.  Counsell et al. extend the testing taxonomy of van Deursen using  the inter-dependencies of the refactoring types \cite{Counsell2006}. Pipka proposed the Test-first Refactoring approach for adapting unit tests according to software changes  \cite{Pipka2002}. 

All the previous does not provide a clear description on how to refactor the tests in a safe manner. They lack an
 operative manner for verifying if test code refactoring modifies its external behavior. 
In our work we address this shortcoming proposing mutation testing as safety net for test code refactoring. We describe how to use
mutation coverage to obtain an operative evaluation of the  behavior preservation of the refactored test.

\section{Conclusion and Future Work}
\label{Conclusion}
Test code refactoring is an important maintenance activity performed to keep it synchronized with production code and avoid its quality erosion \cite{Rompaey2006}. 
For the test code has been identified \textit{ad hoc} refactoring types and peculiar design smells \cite{Deursen2001}. 
Nowadays, refactoring of the test code is riskier than the one performed on the production code. Indeed, the latter benefits from the safety net
provided by the test suite. On the other hand, test code refactoring does not have an equivalent safeguard to assure that external behavior of the test code is preserved.

In this paper, we propose mutation testing as a safety net for test code refactoring.  By conducting the empirical experiments on two projects, we show that mutation testing is (1) suitable for identifying a change on the external behavior of a refactored test and (2) can be used to identify which part of the test code was improperly refactored. However, our approach is limited by the fact that the
refactoring must be restricted to the test code. Any change to production code would result in a different set of mutants which makes the comparison between two versions much harder. However, the developer can avoid this problem by doing the refactoring in two separate phases; First on the production code, and then on the test code. In this case, the behavioral change of the production code
 can be detected using the test suite, and then our method can be still applied.%
 It is worth noting that the reliability of this process to detect behavior changes depends on the accuracy of mutation testing which, in turn, depends on the type of mutation operators that are used. A different set of mutation operators would lead to different mutants being generated, resulting in a different detection ability for the process.

In this empirical study, we take into account a small open source project. 
In the future, we plan to extend this analysis to several other projects with a modified test code. 
In particular, we will investigate projects where the common test refactorings are performed \cite{Deursen2001, Meszaros2006}. There is a lack of empirical studies on evaluation of the proposed test code smells in different different setups (e.g. industrial settings). This can be investigated alongside our method of detecting improper refactorings in these setups, quantifying the probability of occurrence of such code smells, and assessing the risks of refactoring the code to eliminate such smells.
 In addition, we plan to create a dataset with seeded improper refactorings that can be used as a test bench.

\needspace{3\baselineskip}
\section*{Acknowledgment}
This work is sponsored by the Institute for the Promotion of Innovation through Science and Technology in Flanders through a  project entitled Change-centric Quality Assurance (CHAQ) with number 120028.
\balance

\end{document}